\documentclass[prd,preprint,aps,amsmath,amssymb,showpacs,nofootinbib]{revtex4-2}

\usepackage{graphicx} 
\usepackage{dcolumn}  
\usepackage{bm}
\usepackage{hyperref}
\usepackage{color}
\usepackage[normalem]{ulem}
\usepackage{comment}
\usepackage{float}

\usepackage{ulem}

\begin{document}

\preprint{PITT-PACC-2119}

\title{Amplifying the Chirp: Using Deep Learning (U-Nets) to filter signal from noise in LIGO data}

\author{Akshay Ghalsasi}
\email{akg53@pitt.edu}

\affiliation{%
Pittsburgh Particle Physics, Astrophysics, and Cosmology Center, Department of Physics and Astronomy,
University of Pittsburgh, Pittsburgh, USA
}%

\date{\today}

\begin{abstract}
The direct detection of gravitational waves by LIGO has heralded a new era for astronomy and physics. Typically the gravitational waves observed by LIGO are dominated by noise. In this work we use Deep Convolutional Neural Networks (specifically U-Nets) to filter a clean signal from noisy data. We present two realizations of U-Net filters, the Noise2Clean U-Net filter which is trained using noisy and clean realizations of the same signal, as well as Noise2Noise U-Net which is trained on two separate noisy realization of the same signal. We find that the U-Nets successfully filter signal from noise. We also benchmark the performance of U-Nets by using them to detect the binary presence or absence of gravitational wave signals in data.
\end{abstract}

\maketitle

\newpage

\section{Introduction}

The observation of gravitational waves by Laser Interferometer Gravitational-wave Observatory (LIGO) \cite{LIGOScientific:2014pky} has opened up an entirely new observational window, with far reaching implications for astrophysics, particle physics and cosmology. The LIGO-Virgo-Kagra (LVK) gravitational observatory \cite{LIGOScientific:2014pky,VIRGO:2014yos} has detected $\mathcal{O}(100)$ events so far, with the O4 run expected to detect several hundred more events.

The events observed in LIGO have a very small amplitude compared to noise. In order to reliably detect a signal, the process of matched filtering is used (see for e.g. \cite{Schmitz:2020syl}), where the measured data is compared against a template bank of simulated signals. This process is expensive since the template bank contains $250,000$ different signals, corresponding to different parameters of the merger i.e. black hole masses, inclination of the binary etc.

 Matched filtering provides valuable information regarding the merging binary system. In this work, we will focus on a simpler task, i.e. detection of presence or absence of gravitational waves in a given measured time series. We take our inspiration and data from the Kaggle G2Net competition which focuses on the same task \cite{KaggleG2Net}. We will benchmark our answers to the dataset provided in the Kaggle competition (thus henceforth we will refer to this as the benchmark dataset). The metric for binary classification in the Kaggle competition is the area under the curve (AUC) for our receiver operating characteristic (ROC) curve, and we will use the same here.

Increasing compute and advances in deep learning has resulted in deep neural networks being used to solve a variety of previously tough problems. Deep learning has also been previously used for detecting signal in LIGO observations as well as parameter estimation \cite{George:2017pmj,Yan:2021wml}. The signal observed at LIGO can be converted into the so-called ``chirp plot'' by performing a continuous wavelet transformation (CWT) of the signal. The result of the CWT transformation is an image which shows the presence of frequencies as time progresses (see Fig. \ref{fig:whitening} for an example). In the presence of a gravitational wave signal shows the presence of a ``chirp'' i.e. a signal with increasing frequency as time progresses. This makes gravitational waves a particularly nice targets for convolutional neural networks, especially deep vision models which excel at the task  of parsing a given image. This is the strategy we will use in this work, with the added innovation of filtering out the noise and making it easy for the vision CNN to detect the  presence of the gravitational wave.

The remaining paper will be divided as follows. In  Sec. \ref{sec:data}, we will describe the data we use as well as the basic preprocessing we perform to enhance the signal compared to noise. Here we will also describe the synthetic data we generate to train our Noise2Clean U-Net. In Sec. \ref{sec:base} we establish a baseline against which we will benchmark the effects of our U-Net filters. In Sec. \ref{sec:U-Net} we will describe the architecture of the convolutional neural network U-Net and our modification of it to filter the signal from the noise. In Sec.\ref{sec:N2C} we will describe the training of our Noise2Clean U-Net using synthetic data we generate.  In Sec.\ref{sec:N2N}, we will describe an alternate filtering technique of training a U-Net through the Noise2Noise procedure. In Sec. \ref{sec:res}, we will discuss our results by using our procedure on our benchmark dataset provided in the Kaggle competition. We will benchmark our results against the standard procedure of simply using a vision net to classify the presence of a gravitational wave signal. Finally we will conclude in Sec.\ref{sec:con} and discuss the multiple avenues of future work.

\section{Data and Preprocessing}
\label{sec:data}

The data that we will use to benchmark our U-Net filtering procedure will come from the Kaggle competition \cite{KaggleG2Net} (see \href{https://www.kaggle.com/competitions/g2net-gravitational-wave-detection/overview}{here}). The data is of 560,000 noisy time-series with labels, half of which have the presence of a GW signal and half do not.  However to reduce the computational time needed, we will only use 1/4th of the total dataset for our benchmark dataset. The data consists of mock observations from the Hanford and Livingston sites of the LIGO detector as well as the Virgo observatory. Each datapoint is a time series of 2 seconds sampled at a frequency of 2048 Hz for each of the LIGO detector Livingston (L), Hanford (H) and Virgo. For the purposes of this work, we will not consider the signal from the Virgo observatory\footnote{This will result in a sub-optimal classification, however the two LIGO detectors have same or similar noise templates making our job of filtering the noise easier}. We will split the data as $90\%$ for training and $10\%$ for testing (validation).

In the benchmark dataset described above, we do not have access to clean signals. To get around this problem we generate synthetic LIGO signals using the GGWD python package \cite{Gebhard:2019ldz} (see this Github \href{https://github.com/timothygebhard/ggwd}{repo}) for code. GGWD package is a wrapper around LIGO's PyCBC package \cite{Usman:2015kfa} which is used by the LIGO collaboration to generate synthetic gravitational waves from mergers of black holes.

While the benchmark dataset is itself a synthetic dataset, the parameters used to generate the gravitational wave signals in that dataset are not known. The exact shape and amplitude of the gravitational wave signal depends on several parameters, such as the mass of the two black holes, the inclination of their orbit to the detectors and the distance of the merger event.  For of our synthetic dataset we use $M_{1,2} \subset [10,80] M_{\odot}$, a uniform prior on spin between $(0,0.998)$, a signal-to-noise ratio of 10, a sin-angle prior on inclination, and uniform prior for all other parameters. We generate 140,0000 synthetic timeseries  (70,000 each for Hanford and Livingston observatories) of 3 seconds interval each. The benchmark dataset consists of "measurements" over a 2 second interval with the merger showing up in approximately the last half a second. To conform to the length of the measurement as well as the location of the merger, we chop our synthetic waveform into 2 second interval, while ensuring that the merger happens between (1.45, 1.95) seconds. We expect that this provides adequate coverage on the benchmark dataset.

To summarzie, we use two different dataseta
\begin{itemize}
    \item \textbf{Benchmark dataset --} This dataset comes from Kaggle G2Net competition and we use it to benchmark our U-Net filtering procedure as well as to train our Noise2Noise U-Net. This dataset consists of noisy time-series half of which have a gravitational signal present.
    \item \textbf{Synthetic dataset --} This dataset comes is generated by us to have access to clean signal time-series i.e. without time-series without the presence of noise. We use this dataset to train our Noise2Clean U-Net.
\end{itemize}

Typically at the raw noisy timeseries it is impossible to determine by eye, the presence of a gravitational wave signal since the noise completely dominates. Plotting the average power spectral density of the time series, we can see that the timeseries is completely dominated by low frequency noise as well as a few single frequency sources at higher frequency (See Fig. \ref{fig:psd} ). Gravitational waves from mergers are expected to have frequencies between $20-500$ Hz, with the amplitude of the signal growing with higher frequency (the ``chirp'').

\begin{figure}
 \includegraphics[width=0.43\linewidth]{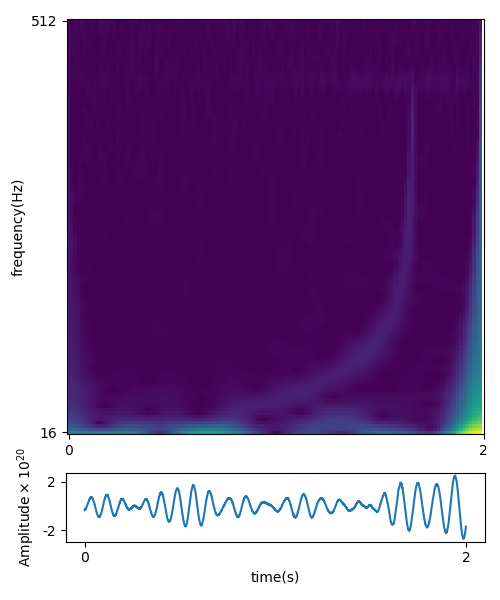}  \quad\quad
 \includegraphics[width=0.43\linewidth]{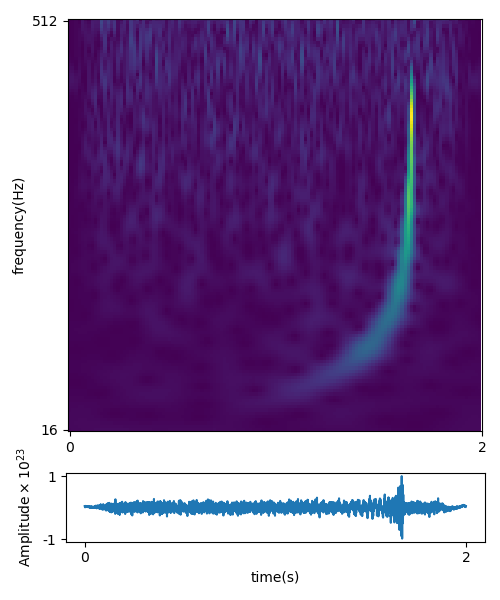} 
 \caption{Raw as well as whitened time series of a particularly prominent signal from our benchmark dataset. Left: The raw gravitational wave time series and its corresponding CWT transform. The signal is not visible to the eye in the time series and is completely overwhelmed by low frequency noise. Right: Whitened time series. The signal is visible in the time series as well as the CWT transform.}
 \label{fig:whitening}
\end{figure}

To make the time series smoothly go to zero at the endpoints, we convolve the filtered signal with a tukey window with $\alpha = 0.2$. To minimize the low frequency noise, we perform the procedure known as whitening. In Fig. \ref{fig:whitening}  we have plotted an example from the raw data from our benchmark dataset. As can be seen, whitening the signal enhances the gravitational wave signal compared to noise. Further details on the whitening procedure have been provided in the Appendix \ref{app:whitening}. The above prepossessing is performed before training in all of our procedures described below.

In all of the work, we do not use the time series signal itself but rather the Complex Wavelet Transform (CWT) of the signal. The CWT is obtained by convolving the time series signal with a wavelet that is continuous in time and frequency space. The output of a CWT of a time series is a 2 dimensional ``image'' which shows the presence of frequencies as a function of time. The CWT transform of a binary BH merger gives rise to a a characteristic ``chirp'' plot (see eg. Fig. \ref{fig:whitening}) where the prevalent frequencies increase with increasing time until the BHs merge. Note that our complex CWT transform has both the real and imaginary part, but we only consider the magnitude here, thus losing the phase information. Most of our work will involve enhancing the chirp with respect to the background noise to make it more prominent and detectable.

\section{Establishing a Baseline}
\label{sec:base}

To quantify the performance of our methods, we need to establish a baseline to compare ourselves to. Here we turn to the standard technique which involves converting the time series signals to an image via a CWT transformation, and then use transfer learning using vision nets to detect the presence of a chirp, thereby classifying the presence or absence of gravitational waves \footnote{This method was the most prevalent in the Kaggle G2Net competition, and performs almost as well as the best solution.}. In particular, we will use EfficientNet architecture \cite{tan2019efficientnet} ( specifically EfficientNetB7)for classifying the presence of gravitational waves.

For our EfficientNetB7, we remove the top layer and replace it with a fully connected binary classifier (a single neuron as output). We will establish two benchmarks, the first one where only the last layer of EfficientNet is trained and all the other layers are frozen, and one where we train the entire EfficientNet. The EfficientNet is initialized with ImageNet weights. Our training pipeline is shown in Fig. See Table. \ref{tab:results} for baseline benchmark results.

The EfficientNet takes an input of three channels (RGB). The U-Net filtered CWT images of Hanford and Livingston serve as two of the three channels. The third channel is an ``average'' constructed out of the the Hanford and Livingston CWT images defined as $2\frac{H*L}{max(H+L)}$, where $*$ refers to an elementwise multiplication.

\begin{figure}
 \includegraphics[width=0.9\linewidth]{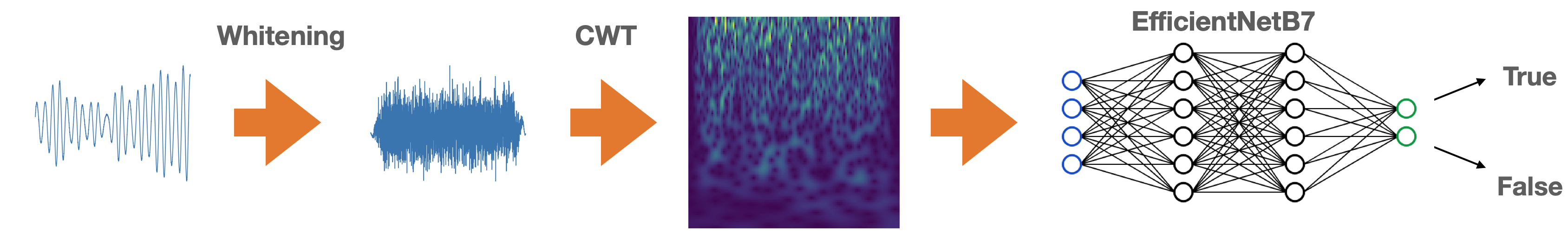} 
 \caption{Training pipeline to establish our benchmark.}
 \label{fig:baseline}
\end{figure}


\section{U-Net as a filter}
\label{sec:U-Net}

Even with the data preprocessing described in Section \ref{sec:data}, the limitation on achieving a perfect classification of the presence of gravitational wave signals comes from the presence of noise which in most cases overwhelms the signal. While the EfiicientNet does perform the task of detecting features within images, and is effectively ``denoising'' the image by focusing on features that indicate the presence of a gravitational wave signal, we want to investigate whether better classification can be obtained if we present the EfficientNet classifier with a denoised image.

In this work, we will use U-Nets to filter noise from signal. U-Nets were originally invented for the purposes of biomedical image segmentation to detect the presence of tumors in medical images \cite{ronneberger2015u}. They are also used for image segmentation in autonomous driving. Since they segment images into relevant and irrelevant parts, they are a natural choice for filtering signal from noise in an image. The U-Net architecture we use is described in more detain in the Appendix \ref{app:U-Net}.

Since the input and output dimensions of a U-Net are the same we can take two approaches to filter signal from noise. The first one involves training a noisy GW signal as input and the corresponding clean GW signal as the target. Then we can train the neural network to filter out the signal from the noise. This is commonly referred to as Noise2Clean (N2C) in literature. The second approach involves training the U-Net where both input and output are separate noisy realizations of the same signal. This method is known as Noise2Noise (N2N) \cite{lehtinen2018noise2noise}. The trained U-Net then learns the statistical average of noisy signals which is a clean signal. We will implement both N2C and N2N neureal networks to filter noise from signal. 

\subsection{U-Net Noise2Clean Filter}
\label{sec:N2C}

The image from the CWT transformation of the clean signal from our synthetic dataset acts as the target for training of our neural network \footnote{We add a small amount of Gaussian random noise to the clean signal. This helps the neural network to not get stuck in the false trivial minima where it simply convolves the input image to zero.}. For input, we use the CWT image of the noisy (signal+noise) signal  measurement. The signal is the synthetic signal we generated, whereas the noise part of the measurement is sampled randomly from the pure noise (no gravitational wave present) data of the benchmark dataset. The noisy measurement is then preprocessed by the prescrpition given in Sec. \ref{sec:data} and the CWT of the processed noisy signal is the used as the input in the U-Net training. The U-Net is the trained by iterating over the signals (130,000 training, 10,000 validation) in the synthetic dataset for several epochs, until the relevant metric (`Validation RMSE') converges. Each time the metric converges, we lower the SNR of the signal by multiplying it with a $\mathcal{O}(1)$ number and retrain. This allows us to filter even weak signals. This process is terminated once we confirm that a large part of the validation data-set is no longer being filtered appropriately.

\begin{figure}
 \includegraphics[width=1.\linewidth]{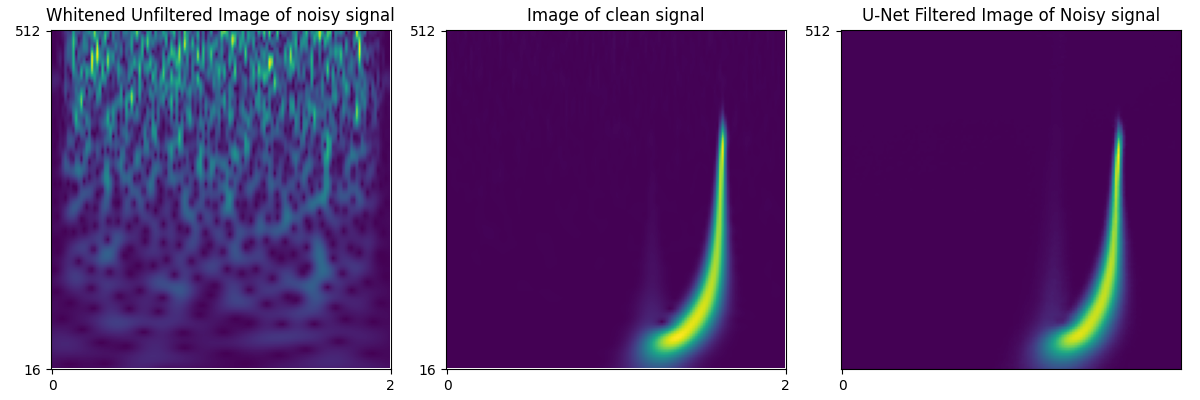}
 \caption{In the figure above we show the effects of our N2C U-Net filter on the validation dataset. The left figure (the input to the U-Net) shows the CWT image of a noisy signal from the validation data-set. The signal is not prominently visible through the noise. Middle figure (the target of the U-Net) is the CWT image of the true clean signal. The right figure (the output of the U-Net) shows the effect of the N2C U-Net applied to the noisy CWT image. As can be seen it reconstructs the true signal reliably. }
 \label{fig:n2c}
\end{figure}

\begin{figure}[H]
 \includegraphics[width=1.\linewidth]{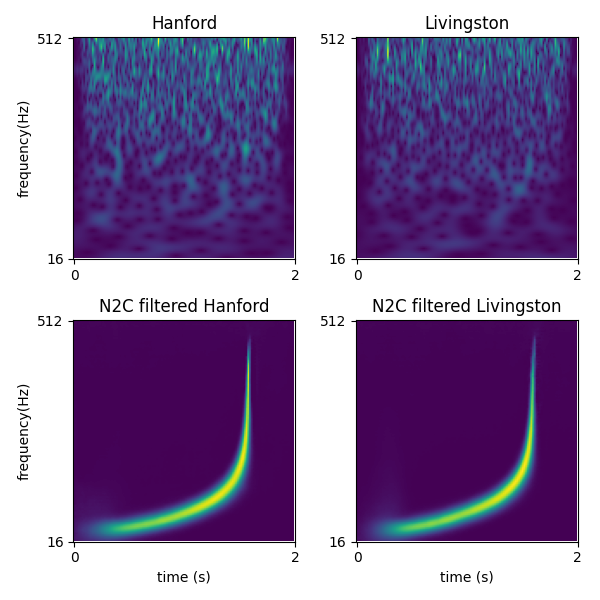}
 \caption{In the figure above we show the effects of our N2C U-Net filter on the validation data-set. The top figures (the input to the N2C U-Net) shows the  CWT image of a whitened noisy signal from our benchmark validation dataset. The signal is not prominently visible through the noise. The bottom figures are the corresponding N2N filtered images. As can be seen the N2N U-Net succeeds in filtering out the signal ``chirp'', even thought its not visible in the noisy images.  }
 \label{fig:n2c-bm}
\end{figure}

In Fig. \ref{fig:n2c} we have plotted the trained U-Net filter applied to the validation dataset of the synthetic GW dataset we generated. It can be seen that the U-Net filters out the signal that is not obviously visible in the CWT images. In Fig. \ref{fig:n2c-bm}, we show the effect of the U-Net N2C filter on our benchmark dataset. As can be seen the U-Net generalizes well to signal not generated by us and successfully filters the signal for our benchmark dataset.

\begin{figure}
 \includegraphics[width=1.\linewidth]{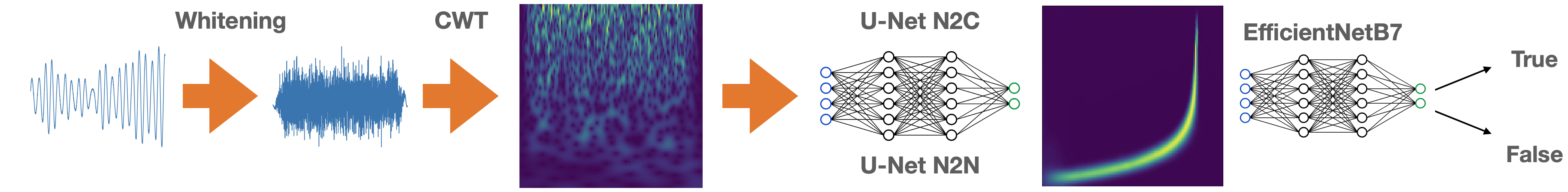}
 \caption{The training pipeline we use to benchmark with the presence of the U-Net filter. The U-Net seeks to filter the signal from noise potentially making the detection of gravitational wave easier. }
 \label{fig:U-Net-n2c-effnet}
\end{figure}

In order to quantify the effect of the U-Net in classification of the presence of gravitational waves, we add the U-Net to our classification pipeline, by filtering the CWT images of the noisy signals from the benchmark dataset through our trained U-Net before feeding it to the EfficientNet for classification (See Fig. \ref{fig:U-Net-n2c-effnet}). Since we train the filter only on the presence of a signal, we will be prone to false positives. As mentioned in the previous section, the three channels of the input to the EfficientNet are the U-Net filtered images of Hanford (H) Livingston (L) and the average $\frac{H*L}{max(H+L)}$. Since the gravitational wave signals at Hanford and Livingston signals will overlap in the U-Net filtered image, the averaging procedure we perform seeks to remove any false positives which would not necessarily overlap. We train the EfficientNet for the same benchmarks as our baseline model. Our results are reported in Table \ref{tab:results}

\subsection{U-Net Noise2Noise}
\label{sec:N2N}

Given two separate noisy realizations of the same signal, if the noise has zero mean, one can train a U-Net using one noisy realization as input and the other as target. Since noise in uncorrelated while the signal is correlated, the network learns to predict the average of the noisy realizations, which is the clean signal. This method of filtering a noisy signal is called Noise2Noise \cite{lehtinen2018noise2noise} since it is based on training on a pair of noisy signals.

Given the fact that LIGO has two detectors (Hanford and Livingston) with similar noise profiles, the Noise2Noise method seems at first like an ideal approach to filter the noisy data. However, though Hanford and Livingston are observing the same merger event, the time of arrival of the signal at Hanford and Livingston can be as different as 10 ms.  Thus although the low frequency part of the signal will still be reasonably correlated in the time series, the high frequency part, corresponding to the actual merger will not be correlated and can be significantly out of phase. However the CWT transform trades accuracy in frequency for accuracy in time, thus the appearance of the signal in the CWT transform image of the Hanford and Livingston signal will overlap reasonably well and can be used in training a Noise2Noise (N2N) filter.

\begin{figure}[H]
 \includegraphics[width=1.\linewidth]{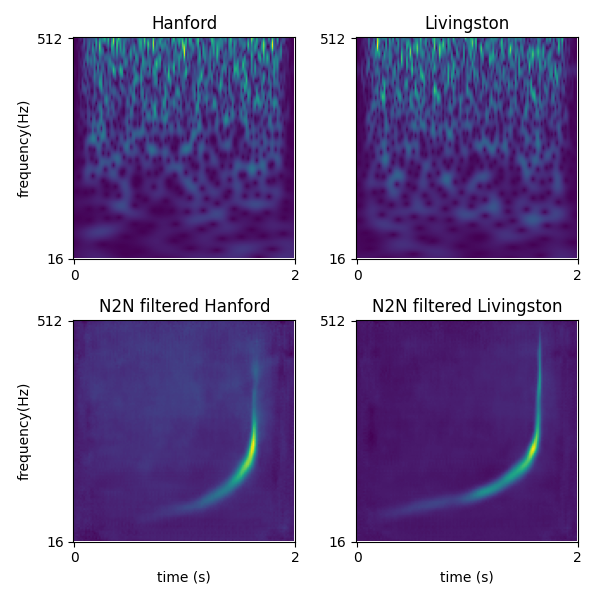}
 \caption{In the figure above we show the effects of our N2N U-Net filter on the validation data-set. The top figures (the input and target to the N2N U-Net) shows the  CWT image of a whitened noisy signal from our benchmark validation dataset. The signal is not prominently visible through the noise. Bottom figures are the corresponding N2N filtered images. As can be seen the N2N U-Net succeeds in filtering out the signal ``chirp'', even thought its not visible in the noisy images. }
 \label{fig:n2n}
\end{figure}

The image of the CWT transform of the time series only carries information of the amplitude of the signal and the phase is ignored. Thus our Noise2Noise images do not contain noise that has zero mean and the filter will learn to project the noise from one image to another thus learning the ``average" of the CWT of the noise in the process. To get rid of the noise we train two Noise2Noise filters. The first one is trained purely on data with the presence of a noisy signal and the other is trained purely on noise only data. Thus applying both the filters and taking their difference should effectively subtract the noise revealing the presence (or absence) of a signal. Since we already have access to two noisy realizations of the same signal (Hanford and Livingston) in our benchmark dataset, we do not need to generate a synthetic dataset to train our U-Net N2N filter, we simply train using our entire benchmark dataset (i.e. we don't restrict ourselves to 1/4th of subset of the data we use for benchmarking). To augment our training we randomly swap our input and target to be the Hanford and Livingston signal. It is important to note that while training a Noise2Noise U-Net, initially the metric of interest ('Validation RMSE') will converge within two epochs and then remain constant. However that does not mean the Noise2Noise U-Net is no longer being trained. In fact it is essential to keep training for a large number of epochs though the training metric does not change. Here we train our Noise2Noise U-Net for 10 epochs. Below we demonstrate the filter effect of our U-Net N2N filter on our benchmark validation dataset (see Fig. \ref{fig:n2n}). As is the case with the N2C filter, we then benchmark the effect of adding our filter to the training pipeline as demonstrated in Fig \ref{fig:U-Net-n2c-effnet}. The results are presented in Table \ref{tab:results}.
\\
$~~~~~~~$A few words about the Noise2Noise filtering approach are in order. The real power of this approach lies in the fact that we did not need to generate a mock dataset of binary mergers in order to train our U-Net filter. In this sense a N2N filtering approach is completely data driven (apart from knowing the binary presence or absence of signal in the dataset). Thus this approach can be in principle be used on real LIGO data to detect signals for which a template is not known. With more LIGO detectors coming online, our ability to pair input and target in our U-Net grows combinatorially, presumably increasing the effectiveness of the N2N filtering method (although further study is required to quantify this statement).

\section{Results}
\label{sec:res}

We tabulate the results of our experiment to quantify the effect of our U-Net filter \ref{tab:results}.
\begin{table*}
\begin{center}
\begin{tabular}{ |c|c|c|c| } 
 \hline
 Model & EfficientNet Size & Validation AUC & Validation Accuracy \\ 
 \hline
 EfficientNet & Last Layer Only & 0.7355 & 0.6788\\ 
 EfficientNet & All Layers & 0.8344 & 0.7657\\ 
 \hline
 U-Net N2C + EfficientNet & Last Layer Only & 0.8206 & 0.7535\\ 
 U-Net N2C + EfficientNet & All Layers & 0.8329 & 0.7577 \\ 
 \hline
  U-Net N2N + EfficientNet & Last Layer Only & 0.8086  & 0.7497 \\ 
 U-Net N2N + EfficientNet & All Layers & 0.8209  &  0.7604 \\ 
 \hline
\end{tabular}
\end{center}
\caption{Results of our experiments. The model refers to the pipeline we use for training our classifier. The EfficientNet Size refers to the size of the EfficientNet we use in training (i.e. train only the last layer vs. all of the EfficientNet). We report the validation AUC and accuracy of our model. } \label{tab:results}
\end{table*}

First we train our benchmark dataset on the last fully connected layer of EfficientNet. The worst  performance comes from only training the last fully connected layer of our EfficientNet. This is to be expected since this includes only a few trainable parameters. Moreover the ImageNet weights we use to intialize our EfficientNet and which we keep frozen during our training are trained on everyday images which looks very different than the ``chirp'' of the gravitational wave signal. Thus since the initial layers don't do a good job on extracting features relevant to the presence of a gravitational waves, training the last layer only adds limited value to our binary classification. This can be seem in the small validation AUC and accuracy in Table \ref{tab:results}. 

This is mostly remedied by applying a U-Net filter to our CWT images of the gravitational wave signals. Both U-Net N2C and N2N manage to clean up the signal well as well as evidenced by the leap in the AUC after training the last layer of the EfficientNet on the filtered images. This is confirmation of the results we found in Secs.~ \ref{sec:N2C}, \ref{sec:N2N}.

While the U-Net parameters are frozen during the training the EfficientNet, we have already trained the U-Net parameters to extract a signal if present. In priciple we can replace the EfficientNet that we use for classification after U-Net filtering with a custom CNN architecture. This is because the U-Net filtered images have a pretty simple representation if a signal is present. However, to keep comparisons valid we stick to using EfficientNet for classification.

While the above results are encouraging they are also expected since we are effectively adding more trainable parameters in our classification problem. To really see if our U-Net filters are performing beyond what's expected we need to ask if the U-Net can filter or amplify signal to a greater extent than what the EfficientNet itself can do. Towards this end we unfreeze all the layers of EfficientNet and train the CWT images of our benchmark dataset on the whole of EfficientNet itself.

Training the whole of EfficientNet is training the network to extract features relevant to the presence (or absence) of a GW signal in the noisy data. Thus the EfficientNet is itself performing the job of ``filtering'' the signal. However it's  unlikely it's doing so in a physics interpretable way the U-Nets filter signal. 

Training all the layers of the EfficientNet gives the best score in our metric of choice  i.e validation AUC. Adding either a U-Net N2C or N2N filter to preprocess the images before they are used to train all of the EfficientNet seems to decrease the validation AUC as well as accuracy slightly. Although the differences between using and not using a U-Net to filter our noisy CWT images is very small (less than 1/5th of the percentage points), this result is surprising if not a little disappointing since we had hoped the U-Nets would enhance the presence of signals the EfficientNet cannot detect by itself.

\begin{figure}
 \includegraphics[width=0.45\linewidth]{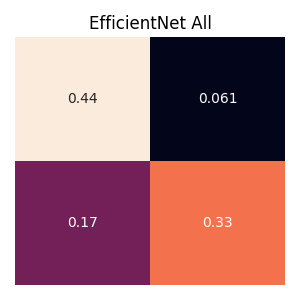}
 \includegraphics[width=0.45\linewidth]{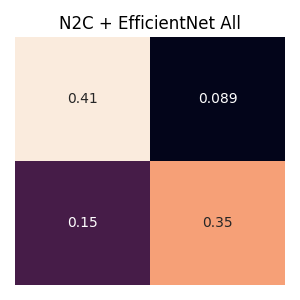}
 \caption{The confusion matrix using all of the EfficientNet and using a N2C filter before using the EfficientNet. Clockwise the confusion matrix reports fractions of true negative, false positives, true positive and false negatives. As can be seen using the N2C filter increases false positives while reducing false negatives.}
 \label{fig:cm}
\end{figure}

To further diagnose the issue with our U-Net filtering method we try to ascertain if the U-Nets are missing very weak signals (that the EfficientNet  architecture can in fact detect by itself) i.e. false negatives, or our U-Net filtering method is hallucinating a signal where none is present i.e. false positives. Looking at the confusion matrices (using a threshold of 0.5) shown in Fig. \ref{fig:cm} it is evident that using a N2C filter before using the EfficientNet increases the false positives while reducing the false negatives \footnote{We choose U-Net N2C for plotting the confusion matrix since it performs better.}. Thus it seems that our method of U-Net filtering introduces additional false positives (while reducing false negatives) despite our efforts to reduce them by making sure that the filtered ``signals'' coincide (which they wouldn't necessarily if the filter hallucinated a signal). In Sec.~\ref{sec:con} we will briefly discuss strategies to mitigate false positives as part of our future work.

As can be seen the U-Net N2N model performs the worst. Some of this is expected since we train the Noise2Noise U-Net on noisy data. However the score of U-Net N2N filter is limited by the number of training samples and can be improved by training on a larger dataset. Moreover the Noise2Noise strategy is expected to give better results if the phase information of the CWT transform is included in our training. We discuss this further in Sec. \ref{sec:con}

\section{Conclusions and Future Work}
\label{sec:con}

In this work we have shown that U-Nets can be used to filter a clean gravitational wave signal from noisy data. We explored two ways of constructing a U-Net. The first method relied on using a Noise2Clean approach where we used noisy data as input and clean data as target and trained the U-Net to filter out the noise. For this method we had to create synthetic gravitational wave signals to train our network and hoped it generalized well to the benchmark dataset we used (which it did to a large extent). Our second approach (Noise2Noise) was to use to separate noisy realizations of the same signal. This was easily accomplished since LIGO has two detectors with a similar noise PSD (i.e. Hanford and Livingston). Although the signals are not coincident in time in the two detectors, their CWT transformed images are coincident enough for us to use this approach. We used two U-Nets, one trained on positive samples only and the other on negative samples only.  In this approach, there was no need generate mock dataset, instead we used our benchmark dataset from the Kaggle G2Net competition. Since we did not need access to clean timeseries the N2N approach is more data-driven than the N2C approach.

We found that the U-Net succeeded in filtering signal from the noise. However filtering the signal with U-Nets produces too many false positives. Additional detectors would certainly help with the problem of false positives. Additionally including the phase information should also help in establishing a coincidence between the two signals. If including the phase information we are able to reconstruct all or part of the merger in one detector, then we can use that waveform to perform matched filtering on the other detector(s) to make sure the signal appears in other detectors as well. These and other standard strategies to mitigate overfitting and hence false positives will be left for upcoming work.

We also propose the possibility of using the U-Nets as a trigger for detecting the presence of gravitational waves (since once trained, inference using the U-Nets will be quick). If the waveform time-series can be constructed phase information, we can use this to provide an initial guess for the matched filter thus speeding up the matched filtering procedure. Further investigation along these lines is needed and we leave that for future work.

The fundamental shortcoming of our approach was that we discarded the phase information of the signal and only trained using the amplitude of the complex CWT transform. Keeping the phase information will allow us to reconstruct the time series of the clean signal by performing an inverse CWT transform. Doing this will allow us to define a loss function by comparing the time series that the U-Net outputs with the expected one and might lead to better filtering. In fact such an approach was used for speech enhancement in noisy audio \cite{kashyap2021speech} by implementing a Deep Complex U-Net \cite{choi2018phase}. We leave this implementation for our upcoming work.

The U-Net N2N method can be used to extract a signal when no template is available. Thus it can be used to detect more exotic signals than mergers. If the binary presence of a signal is known i.e. the time series is not consistent with noise only in all of LIGO detectors, then such a dataset can be used to find the new exotic signal. Thus we expect an approach that uses anomaly detection (to detect the binary presence of a signal) combined with a N2N U-Net can reveal exotic sources if they are present. On the other hand, if a signal is always present such as a continuous gravitational wave signal, the U-Net N2N approach can be used to enhance such a signal.

The U-Net N2N approach can also be used in other astrophysical data where a signal with unknown shape is known to be present but buried under noise. An example would be pulsar observations where the pulse shape and hence the position (exact time of arrival) is not known. One can use the Noise2Noise approach on pulses within a window of observations where the pulsar properties are not expected to change significantly. Whether this approach provides benefit over the standard stacking/folding procedure to will need to be investigated.

\section{Acknowledgements}
The work of A.G. is supported by the U.S. Department of Energy under grant No. DE–SC0007914. We would like to thank the Aspen Center for Physics where part of this work was completed. We would also like to thank Ben Michel, Jared Evans and  David Shih for helpful discussions. We would also like to thank Brian Batell, Matthew Low and Jared Evans for providing comments on the draft.

\appendix
\section{Whitening}
\label{app:whitening}

\begin{figure}
 \includegraphics[width=1.\linewidth]{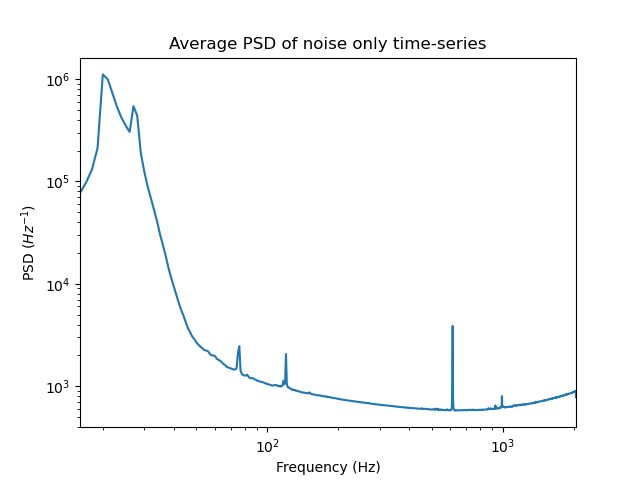}
 \caption{The averaged PSD of the normalized time series of noise only data.}
 \label{fig:psd}
\end{figure}

Typically the power spectrum of a time series, i.e. the contribution of each frequency to a given signal typically depends on frequency. For our gravitational wave time series the power spectrum density (PSD) of the noise is dominated at low frequencies by seismic noise and at high frequencies by the shot noise of the laser. It also includes certain dominant peak frequencies. The average PSD of the noise only time series are plotted in Fig. \ref{fig:psd}

Whitening as the name suggests is a process that removing the frequency dependence of the PSD (white noise has no frequency dependence). In order to do this we take the FFT of the time series, divide by the PSD and take the inverse Fourier transform. This gives us the whitened time series.

\section{U-Net Architecture}
\label{app:U-Net}

The U-Net architecture consists of a contractive (downsampling) path followed by an expansive (upsampling) path (See Fig.). The basic building block of the contractive path is two successive applications of a convolution consisting of $n$ filters and a rectified linear unit (ReLu) activation. $n$ is chosen as $n_{0} \times 2^{l-1}$, where $l$ is the layer number. We choose $n_{0} = 32$.  We choose padding as `same' for the convolutions. In a standard U-Net this is usually followed by a MaxPooling application. However we replace this with another convolution and a stride of 2. This contractive block is then applied in sequence to the input image for a prescribed number of times (we choose 6). In the final application of the contractive path, we do not apply the final convolutional pooling with stride of two. This creates a the contractive path of the U-Net. At each level of the contractive path (each application of the building block of the contractive path), spatial information is sacrificed in favor of features. 

\begin{figure}
 \includegraphics[width=1.\linewidth]{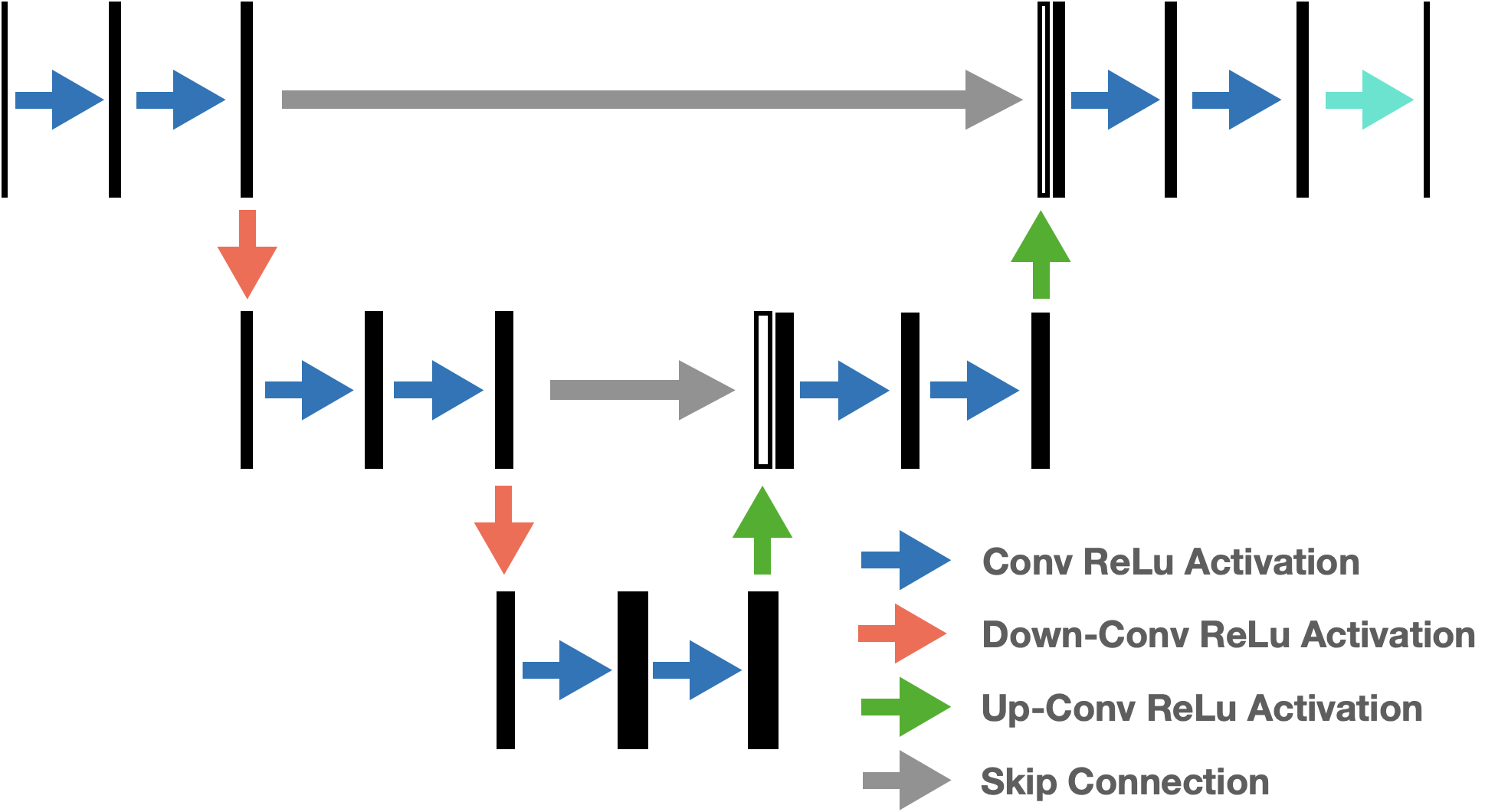}
 \caption{The U-Net architecture. We use a 6 layer U-Net, but have shown only 3 layers for purposes of illustration}
 \label{fig:U-Net}
\end{figure}

This is followed by the expansive path, the building block of which is deconvolution (with $n = n_{0} \times 2^{l-1}$ filters, stride of two, and kernel size of three)  followed by concatenation with a block from the at the same level. This concatenation is done through skip connections between the contractive and expansive path. Thus the upsampling block combines the spatial information from a given level with high-level features from the preceeding deeper level. We then follow with two more convolutions (with $n = n_{0} \times 2^{l-1}$ filters, stride of two, and kernel size of three). This upsampling is repeated for the same number of layers. Finally we perform a convolution with $n = 1$ to recover the same dimension as the input. See Fig. \ref{fig:U-Net} for an illustration of the U-Net.

We then train this U-Net based on our choice of inputs and 
target i.e. N2C or N2N.

\newpage
\bibliographystyle{apsrev4-1}
\bibliography{refs}

\end{document}